\documentclass{article}
\usepackage{spconf,amsmath,graphicx,listings}
\usepackage{booktabs}
\usepackage{pythonhighlight}
\newcommand{\etal}{\textit{et al. }}
\newcommand{\citep}[1]{\cite{#1}}
\newcommand{\citet}[1]{\cite{#1}}

\newcommand\norm[1]{\left\lVert#1\right\rVert}
\title{Onssen: an open-source speech separation and enhancement library}
\name{Zhaoheng Ni\qquad Michael I Mandel}
\address{City University of New York\\
        365 Fifth Avenue, New York, NY 10016}
\begin{document}
%
\maketitle
\begin{abstract}
Speech separation is an essential task for multi-talker speech recognition. Recently many deep learning approaches are proposed and have been constantly refreshing the state-of-the-art performances. The lack of algorithm implementations limits researchers to use the same dataset for comparison. Building a generic platform can benefit researchers by easily implementing novel separation algorithms and comparing them with the existing ones on customized datasets. We introduce "$onssen$": an open-source speech separation and enhancement library. $onssen$ is a library mainly for deep learning separation and enhancement algorithms. It uses LibRosa and NumPy libraries for the feature extraction and PyTorch as the back-end for model training. $onssen$ supports most of the Time-Frequency mask-based separation algorithms (e.g. deep clustering, chimera net, chimera++, and so on) and also supports customized datasets. In this paper, we describe the functionality of modules in $onssen$ and show the algorithms implemented by $onssen$ achieve the same performances as reported in the original papers.
\end{abstract}
\section{Introduction}
\label{sec:intro}

Overlaps of different speakers are very common in real-life conversations. While it is easy for humans to focus on one speaker in noisy multi-talker environments, it is difficult for machines to achieve a compatible performance. The goal of speech separation is to separate the speech of interest from multi-talker recordings so that it can improve the performance of ASR systems. Though many successful deep learning algorithms separate speech from background noise or music, there was limited progress on the talker-independent speech separation task with deep neural networks. It is the label permutation problem which made the multi-talker speech separation much more challenging \citep{yu2017permutation}. To overcome the label permutation problem, a Permutation Invariant Training (PIT) criterion is proposed to train deep neural networks for speech separation \citep{yu2017permutation}. The algorithm first computes the losses for every possible permutation of the model output and mask target pairs, and choose the one with minimum loss values as the true pair to do back-propagation. The PIT criterion significantly improves the separation performance, it can also be applied to any Time-Frequency (T-F) mask-based deep learning algorithms. Instead of predicting the T-F mask for the speaker, Hershey \etal propose a deep neural network called "Deep Clustering" which transforms the T-F bins to embedding vectors \citep{hershey2016deep}. After training, the embeddings from the same speakers are close to each other. Producing T-F masks is effective by applying clustering algorithms. The mask generated by deep clustering is a binary mask, which is not optimal compared with other soft masks (e.g. ideal ratio mask, phase-sensitive mask). In \citep{luo2017deep}, Luo \etal propose a neural network which outputs the embedding vectors and soft masks at two respective layers and name it as "chimera" network. Wang \etal later improve the chimera network by trying alternative loss functions to achieve much better performance (called chimera++ network \citep{wang2018alternative}).

The competition of speech separation algorithms is not ending. Instead, more and more powerful algorithms are proposed and keep refreshing the best performance in recent two years. The experiments \citep{wang2018end, le2019phasebook, wang2018deep} show that only the magnitude information is not enough for speech separation. They predict the phase of clean speech by using chimera++ network with a waveform-level loss function and achieves better results than that of the original chimera++ network. Luo \etal propose an end-to-end speech separation network called "TasNet"\citep{luo2018tasnet} which separates the audio directly. Later, they change the LSTM architecture to fully-convolutional networks (conv-TasNet) and achieves much better performances \citep{luo2018convtasnet}. \citep{shi2019furcanext} outperforms conv-TasNet by applying an end-to-end dynamic gated dilated temporal convolutional networks called "FurcaNeXt". Liu \etal apply a deep computational auditory scene analysis (CASA) approach and apply a frame-level PIT criterion to generate the masks. The model achieves comparable performance than FurcaNeXt with much fewer parameters. Just recently, Luo \etal replace the 1-D convolutional layers with proposed dual-path RNN layers (DPRNN) \citep{luo2019dual} in the conv-TasNet and again refreshed the state-of-the-art performance on the wsj0-2mix dataset \citep{hershey2016deep}.

It is very exciting to see the fast iterations of deep learning approaches to crack the speech separation problem. On the other hand, the lack of implementations of those algorithms makes it difficult for researchers to compare. Some researchers choose to use the same dataset (wsj0-2mix) and compare their SDR metric scores with the ones reported in the previous papers. However, the dataset is not guaranteed to be the most generalized one. If researchers want to compare the algorithms on a different dataset, they will suffer from re-implementing all algorithms and adapt the feature generation scripts to the new dataset. $nussl$ \citep{nussl} is proposed as an open-source toolkit for music and speech separation. However, in terms of deep learning approaches, it only contains the Deep Clustering network. The training script is also not included in the repository, which makes it difficult to reproduce the result. To overcome this problem for every researcher doing speech separation, we want to build a framework that can train the speech separation models from scratch and give researchers much more flexibility to customize the separation algorithms and the dataset. By forcing all the scripts to follow the unified format, $onssen$ can help people write their novel model architectures and feature generators without much effort.

Section \ref{sec:org} introduces the organizations of $onssen$ modules. Section \ref{sec:baselines} reports the baseline performances of the separation algorithms implemented by $onssen$. We discuss the future development plan of $onssen$ and make the conclusion in Section \ref{sec:conclusion}. More information can be obtained at $onssen$’s online repository\footnote{https://github.com/speechLabBcCuny/onssen}.

\section{Library Organization}
\label{sec:org}
\begin{figure}[h]
        \centering
        \includegraphics[width=0.5\textwidth]{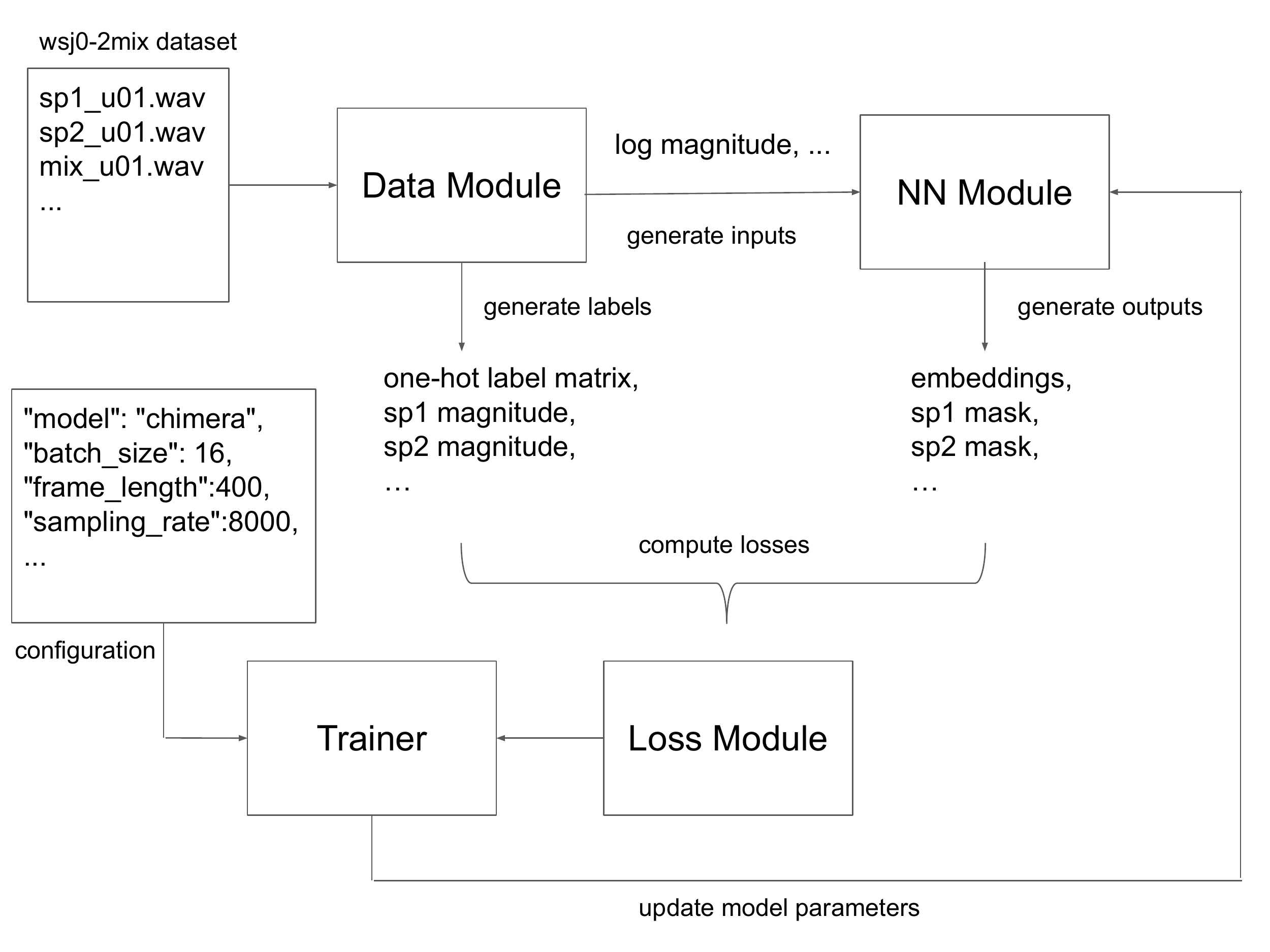}
        \caption{The diagram of $onssen$ training workflow}
        \label{fig:diagram}
    \end{figure}

Figure~\ref{fig:diagram} shows the workflow diagram of the training process. The trainer module reads the configuration JSON file and initializes the NN module, Data module, and Loss module respectively. The data module generates batches of input features for the NN module and the labels for the Loss module. The Loss module takes the model outputs and labels to compute the gradients. Then the trainer updates the model based on the gradients. Hence users can train a customized model by easily adding a configuration JSON to the library without writing the code for feature extraction, model implementation, or training models.
    
\subsection{Data Module}
\label{sec:data_module}
The DataLoader in PyTorch framework is an efficient method to generate training examples. Hence we use it as the basic class for our data module. After initialization, the module returns a DataLoader object which can iteratively generate training batches. The module requires feature\_options from the configuration file. We use wsj0-2mix as an example, the feature\_options contains:
\begin{itemize}
    \item data\_path: the path of the wsj0-2mix root directory
    \item batch\_size: the batch size for the training
    \item frame\_length: the number of frames for each training example. In \cite{hershey2016deep} the length is set to 100. Later experiments \cite{yu2017permutation, wang2018alternative} suggest using longer frame length to improve the separation result. The default value is 400.
    \item window\_size: the window size to generate the Short-Time Fourier Transform (STFT). It is 256 by default for 8 kHz audios.
    \item hop\_size: the hop size of shifting the window. It is set to 64 by default.
    \item db\_threshold: the threshold for detecting the silence regions in deep clustering (will be explained in Section \ref{sec:loss_module}). By default, it is set to -20. 
\end{itemize}

So far we implemented the data modules for wsj0-2mix and Edinburg-TTS datasets \citep{valentini2017noisy}. More modules can be included as long as there exist clean references for all the sources in the mixture. We apply LibRosa \citep{mcfee2015librosa} and NumPy \citep{van2011numpy} libraries for audio processing and feature extraction (e.g. STFT, Mel-spectrogram, magnitude, and so on). The advantage is that all methods are well packaged, which avoids the effort of re-implementation. The disadvantage is both libraries currently don't support operating on GPU, hence there will be a data transition time if the model training happens on GPU. Recently PyTorch releases "torchaudio" which contains many useful audio processing methods. This could be an option for future development.

Since different models require different numbers of input features or labels, the model name is also an argument to construct a DataLoader object. The data module will generate corresponding features and labels based on the model. It is confusing how to assign the input features to the model and assign the outputs and labels to the loss function. To avoid this problem, we force the DataLoader only to generate two objects: "inputs" and "labels". Both are lists of tensors. We also force the model to generate one list of output tensors called "outputs". In this way, all the compatible Data modules and NN modules can be applied for training. We also add essential assertions in the NN modules and Loss modules to make sure the number of arguments and the shape of passed tensors are as expected.

\subsection{NN Module}
All implemented algorithms are stored in the NN module. Each model class is inherited from PyTorch nn.Module class. Every model accepts only one argument "inputs" which is a list of tensors. Each model is required to assert the number of the tensors is as expected. Here we show the code example of the uPIT-LSTM network which predicts T-F masks by giving the log magnitude of mixture speech.
\begin{python}
class uPIT_LSTM(nn.Module):
    def __init__(
        self,
        input_dim,
        output_dim,
        hidden_dim=300,
        num_layers=3,
        dropout=0.3,
        num_speaker=2,
    ):
        super(uPIT_LSTM, self).__init__()
        self.output_dim = output_dim
        self.num_speaker = num_speaker
        rnn = nn.LSTM(
            input_dim,
            hidden_dim,
            num_layers,
            dropout=dropout,
            bidirectional=True,
            batch_first = True
        )
        fc_mi = nn.Linear(
            hidden_dim * 2, 
            output_dim * num_speaker
        )
        self.add_module('rnn', rnn)
        self.add_module('fc_mi', fc_mi)

    def forward(self, inputs):
        assert len(inputs) == 1
        x = inputs[0]
        batch_size, frame_length, _ = x.size()
        self.rnn.flatten_parameters()
        rnn_output, hidden = self.rnn(x)
        masks = self.fc_mi(rnn_output)
        masks = torch.sigmoid(masks)
        masks = masks.reshape(
            batch_size, 
            frame_length,
            self.output_dim,
            self.num_speaker
        )
        return [masks]
\end{python}
Besides the separation algorithms, we also implement a spectral-mapping speech enhancement algorithm proposed in \citep{chen2017improving} as a template for speech enhancement algorithms. 
\subsection{Loss Module}
\label{sec:loss_module}
As mentioned in \ref{sec:data_module}, the arguments for the loss functions are always "outputs" and "labels", which are two lists of PyTorch tensors. Adding certain assertions is important to make sure the loss function fits the need. Here we list all loss functions implemented in the Loss module. Based on the model architecture, they can be separated into two categories: "Deep Clustering" losses and "Mask Inference" losses.
\subsubsection{Deep Clustering Losses}
The loss function of deep clustering in \citep{hershey2016deep} is defined as
\begin{eqnarray}
    L_{\text{DC, classic}} &= &\norm{VV^T - YY^T }_F^2\nonumber \\
     & = & \norm{V^TV}_F^2 - 2\norm{V^TY}_F^2 + \norm{Y^TY}_F^2
\end{eqnarray}
where $V$ is a $B \times N \times D$ embedding matrix generated from the deep clustering network. $B$ is the batch size, $N$ is the number of T-F bins in one training example, and $D$ is the dimension of the embedding vector. $Y$ is a $B \times N \times S$ one-hot matrix representing the dominated speaker in the spectrogram. $S$ is the number of speakers in the audio mixture. Note that the batch size dimension must be separated from the T-F dimension, otherwise it doesn't make sense to apply matrix multiplications to the embedding matrix of one speaker and the label of another speaker.

In \citep{wang2018alternative}, it is suggested that removing the loss of silence regions helps improve the training. Hence the formula can be modified as 
 \begin{eqnarray}
     L_{\text{DC, classic, W}} &= & \norm{W^{\frac{1}{2}}(VV^T - YY^T)W^{\frac{1}{2}}}_F^2 \nonumber \\
      & = & \sum_{i,j}{	w_iw_j[\langle v_i, v_j \rangle - \langle y_i, y_j \rangle]^2}
 \end{eqnarray}
 where $W$ is the weighted matrix for the T-F bins. The simple way proposed in \citep{wang2018alternative} is using binary voice activity weights $W_{\textrm{VA}}$ to filter out the silent regions. $W_{\textrm{VA}}  = \textrm{diag}(w)$ is defined as 
 \begin{equation}
     w_i = \max_k[10\log_{10}(\norm{s_{k,i}}^2/\max_{j}\norm{s_{k,j}}^2)>\beta]
 \end{equation}
where $i$, $k$ represent the indices of the T-F bin and speaker respectively. In other words, if the difference between the clean log magnitude and the maximum of the log magnitude in the utterance is not greater than $0.1 \beta$ for all the speakers at T-F bin $i$, the weight at the T-F bin $i$ is 0, otherwise, the weight value is 1. We adopted the loss function from $nussl$ toolkit \citep{nussl} and modified it to be in the unified format as other loss functions.
 
 \subsubsection{Mask Inference Losses}
 Different from traditional speech enhancement losses, the loss for the speech separation requires the PIT criterion to find the local optimal. We apply the utterance-level PIT (uPIT) criterion to all T-F mask-based loss functions by default. The Magnitude Spectral Approximation (MSA) with PIT loss function is defined as:
\begin{equation}
     L_{\text{MI}, \text{MSA}} = \min_{\pi \in P}\sum_{c}{\norm{\hat{M_{c}}\odot \left| X\right| -
     \left| S_{\pi(c)}\right|}_F^{1}}.
\end{equation}
 $\hat{M_{c}}$ is the generated mask for speaker $c$, $X$ is the mixture magnitude, and $S_{\pi(c)}$ is the clean magnitude for permutation $\pi(c)$. \citep{wang2018alternative} shows that using the L1 norm is better than the L2 norm in the MSA loss function. Hence we use the same setting in our implementation.
 
 As shown in \cite{wang2018alternative, wang2018end, le2019phasebook, wang2018deep}, phase information plays an important role in reconstructing the clean speech from the estimated mask. Thus estimating the phase information by using neural networks is a hot topic in speech separation. We implemented the Truncated phase-sensitive spectrum approximation (tPSA) used by the chimera++ network. The loss function is defined as:
 \begin{multline}
     L_{\text{MI}, \text{tPSA}} = \min_{\pi \in P}\sum_{c} \left\| \hat{M_{c}}\odot \left| X\right| \right. \\ \left. -
     T_0^{\left| X\right|}(\left| S_{\pi(c)}\right| \odot \cos (\theta_{X} - \theta_{\pi(c)})) \right\|_F^{1}
\end{multline}

 The loss function of chimera or chimera++ network is the weighted combination of the deep clustering loss and the mask inference loss. The formula is written as:
     \begin{equation}
       L_{\text{CHI}} = \alpha \frac{L_{\text{DC}}}{N} + (1-\alpha)L_{\text{MI}}
    \end{equation}
where $\alpha$ is set as 0.975 by default. It is also possible to set it as a learn-able parameter in the chimera++ network and optimize it in the training process. 

We show one example of the chimera loss function which combines the deep clustering loss function with the MSA loss function.
\begin{python}
def loss_chimera_msa(outputs, labels):
    assert len(outputs) == 3
    assert len(labels) == 4
    [embedding, mask_A, mask_B] = outputs
    [one_hot_label, 
     mag_mix, 
     mag_s1, 
     mag_s2] = labels
    loss_embedding = loss_dc(
        [embedding], 
        [one_hot_label]
    )
    loss_mask = loss_mask_msa(
        [mask_A, mask_B], 
        [mag_mix, mag_s1, mag_s2]
    )
    return loss_embedding*0.975 + loss_mask*0.025
\end{python}

\section{Baselines}
\label{sec:baselines}

To validate the functionality of $onssen$, we train the implemented algorithms on the wsj0-2mix dataset. The sampling rate is 8 kHz. A 129-dimensional log magnitude is used as the feature for training the models. We don't apply multi-stage training (i.e. train the model on short chunks of audio then re-train the model on longer chunks). We use Adam as the optimizer and set the learning rate to be 0.001. We clip the gradients to be in the range of [-1.0, 1.0]. We train the model for 100 epochs if the validation loss keeps decreasing. The training process will stop if the validation loss doesn't decrease for 6 epochs. The separated speech is generated by multiplying the estimated masks with the mixture STFT and applying inverse STFT to reconstruct the waveform signal.
\begin{table}[ht]
    \centering
    \begin{tabular}{lccc}
    \toprule
    Model  & Mask & SDR & Reported \\
    \midrule
    Deep Clustering & N/A & 7.6 & 6.5\\
    uPIT-LSTM  & MSA & 10.0 & 10.0\\
    Chimera &  MSA & 10.5 & -\\
    Chimera++  & tPSA & 11.0 & 10.9\\
    \bottomrule
    \end{tabular}
    \caption{SDR scores of implemented algorithms by $onssen$ and scores reported in the papers}
    \label{tab:chimeraComparison}
\end{table}

Table~\ref{tab:chimeraComparison} shows the performances of the implemented algorithms and the reported scores in the papers. The results show that $onssen$ can achieve comparable performances with the ones in the original papers.

\section{Future Work and Conclusion}
\label{sec:conclusion}
In the future, we plan to make the current algorithms work for 3 or more speaker mixture dataset. As suggested in \citep{liu2019divide}, the frame-level PIT criterion can find a better local optimal compared with utterance-level PIT. We plan to implement it and verify it on the implemented algorithms. Besides those, we plan to include more end-to-end speech separation algorithms to $onssen$, such as TasNet \citep{luo2018tasnet}, conv-TasNet \citep{luo2018convtasnet}, and DPRNN \citep{luo2019dual}. 

In terms of the training efficiency, the current library supports the training on CPU or single GPU. In the future, we will add support for distributed training which allows users to train models on multiple GPUs or machine clusters.

Of course, it is impossible to add all of the deep learning speech separation algorithms to $onssen$ without the help of the research community. We believe $onssen$ provides an easy and user-friendly framework to help researchers implement their ideas without much effort. We also encourage researchers to follow the $onssen$ workflow guideline and add their customized implementations (e.g. feature extraction scripts for the new dataset, or model definition of new deep neural networks) to the library.

\bibliographystyle{IEEEbib}
\bibliography{refs}

\end{document}